\documentclass[conference,a4paper]{IEEEtran}
\usepackage{float}
\usepackage{bbm}
\usepackage{footmisc}

\usepackage{multirow}
\usepackage{amssymb}
\usepackage{amsmath}
\usepackage{mathrsfs}
\usepackage{amsfonts}
\usepackage{graphicx}
\usepackage{tabularx}
\usepackage{epstopdf}
\usepackage{microtype}
\usepackage{stfloats}
\usepackage{graphicx,subfig,color,multirow,amsmath}
\usepackage{cite}
\usepackage[noend]{algpseudocode}
\usepackage{algorithmicx,algorithm}
\usepackage{color}

\newcommand{\HH}{\dagger}

\newtheorem{theorem}{Theorem}

\makeatletter

\newcommand{\Rmnum}[1]{\expandafter\@slowromancap\romannumeral #1@}
\IEEEoverridecommandlockouts
\title{\huge{Energy Minimization for Mobile Edge Computing Networks with Time-Sensitive Constraints}}
\author{\IEEEauthorblockN{Jun-Jie Yu$^\HH$, Han Wang$^\HH$, Mingxiong Zhao$^\star$, Wen-Tao Li, Hui-Qi Bao, Li Yin, and Mi Wu}\vspace{-1em}\\
\IEEEauthorblockA{National Pilot School of Software, Yunnan University, Kunming, China\\
Email: mx\_zhao@ynu.edu.cn}\vspace{-2.7em}
\thanks{$^\HH$Indicates equal contribution. $^\star$M. Zhao is the corresponding author. This work is supported in part by the National Natural Science Foundation of China under Grant 61801418, in part by Yunnan Applied Basic Research Projects 2019FD-129.}}

\begin{document}
\maketitle
\begin{abstract}
  Mobile edge computing (MEC) provides users with a high quality experience (QoE) by placing servers with rich services close to the end users. Compared with local computing, MEC can contribute to energy saving, but results in increased communication latency. In this paper, we jointly optimize task offloading and resource allocation to minimize the energy consumption in an orthogonal frequency division multiple access (OFDMA)-based MEC networks, where the time-sensitive tasks can be processed at both local users and MEC server via partial offloading. Since the optimization variables of the problem are strongly coupled, we first decompose the original problem into two subproblems named as offloading selection ($\mathbf{P_O}$), and subcarriers and computing resource allocation ($\mathbf{P_S}$), and then propose an iterative algorithm to deal with them in a sequence. To be specific, we derive the closed-form solution for $\mathbf{P_O}$, and deal with $\mathbf{P_S}$ by an alternating way in the dual domain due to its NP-hardness. Simulation results demonstrate that the proposed algorithm outperforms the existing schemes.
  \end{abstract}

\section{Introduction}
With the explosive growth of smart terminals in the IoT era, a large number of applications have emerged that have high computation loads and critical latency (e.g., real-time online games, virtual reality). However, due to the limited energy capacity and computation capability on IoT devices, they cannot fully support the computation-intensive and delay-sensitive services \cite{abbas2018mobile}. To find a way out of this dilemma, a novel computing paradigm called Mobile edge computing (MEC) has been envisioned as a promising technology, which integrates cloud computing and mobile network to offer considerable computation resources at network edge to process the offloaded computation intensive or energy-consuming tasks from IoT devices \cite{mu2019joint}. 

As an effective method to liberate IoT devices from computation-intensive computing workloads, MEC can efficiently reduce the energy consumption of IoT devices, and thus has been considered as a promising architecture for the scenarios with energy-constrained IoT devices \cite{dai2018joint,liu2019dynamic}. To be specific, user association was optimized to minimize the total energy consumption of MEC systems in \cite{dai2018joint}. Meanwhile, the authors in \cite{liu2019dynamic} aimed to minimize users' power consumption while trading off the allocated resources for local computation and task offloading. 

However, offloading energy-consuming workloads to MEC servers also invokes extra latency which significantly affects the quality of experience (QoE) of users, and thus cannot be ignored in the system design. As one of the important metrics to measure the performance of MEC network, time sensitivity has triggered more and more research interests and has been investigated in the literature \cite{ren2018latency,cui2018joint,kuang2019partial}. To be specific, the authors investigated the latency-minimization problem in a multi-user time-division multiple access MEC offloading system \cite{ren2018latency}. Moreover, some pioneering works considered the energy consumption minimization with the QoE requirement of time-sensitive computation tasks \cite{cui2018joint,kuang2019partial}. To satisfy user demands of various IoT applications, the authors in \cite{cui2018joint} found a tradeoff between the energy consumption and latency, and formalized the problem into a constrained multi-objective optimization problem. At the same time, the authors in \cite{kuang2019partial} minimized the weighted sum of the execution delay and energy consumption while guaranteeing the transmission power constraint of IoT devices based on partial offloading.
\raggedbottom

To further improve the utilization of radio resources, Orthogonal Frequency-Division Multiple Access (OFDMA) has been widely applied to the MEC system for various objectives in MEC system, such as profit maximization \cite{paymard2019task}, delay minimization \cite{li2017joint}, the energy consumption minimization with or without the requirement of computation latency \cite{khalili2019joint,yang2019energy}, and the energy-latency tradeoff \cite{zhang2017energy}. Specifically, the authors in \cite{paymard2019task} proposed a priority-based task scheduling policy and jointly optimized the computation and communication resource to maximize the profit of mobile network operator while satisfying users' quality of service (QoS). Furthermore, to minimize the maximum delay of each mobile device, the authors in \cite{li2017joint} considered a partial offloading scheme and developed a heuristic algorithm to jointly optimize the subcarrier and power allocation. At the meantime, QoS was introduced to minimize the energy consumption of a multi-cell MEC network in \cite{khalili2019joint}. Moreover, the authors proposed an energy-efficient joint offloading and wireless resource allocation strategy for delay-critical applications to minimize the total (or weighted-sum) energy consumption of the mobile devices \cite{yang2019energy}, and the tradeoff between energy consumption and sensitive latency was further considered to design the energy-aware offloading scheme in \cite{zhang2017energy}.

However, the inspiring works \cite{zhang2017energy,yang2019energy} only investigated binary offloading without the consideration of partial offloading, which can flexibly allocate resources for computation offloading and local computing, and thus achieve better performace \cite{wu2019computation}, although the QoE requirement of time-sensitive computation tasks was taken into account \cite{ zhang2017energy}. Motivated by the aforementioned issues, we consider partial offloading where mobile date can be computed at both local users and MEC server, and investigate an OFDMA-based MEC network in this paper. Our designed framework aims to minimize the total energy consumption of the whole network, where offloading ratio, computation capability and subcarriers are jointly optimized to satisfy the QoE requirement of time-sensitive computation tasks of users.

\section{System model and problem formulation}
We consider an OFDMA-based MEC system with $K$ users and one base-station (BS) integrated with an MEC server to execute the offloaded data of users, and all nodes are equipped with a single antenna. Denote $\mathcal{K} \triangleq \{1,2, \cdots, K\}$ as the set of users, and let  $\mathcal{N}\triangleq \{1,2, \cdots, N\}$ be the index for multiple orthogonal subcarriers, each of which has bandwidth $B$ and can be assigned to only one user. In this system, we assume that user $k$ has a task described by a tuple of four parameters $\{R_k, c_k, \lambda_k, t_k\}$, where $R_{k}$ indicates the amount of input data to be processed, $c_{k}$ reprsesents the number of CPU cycles for computing 1-bit of input data, $\lambda_k\in[0,1]$ is the proportion of $R_k$ offloading to MEC, while the rest $(1-\lambda_k)R_k$ bits are processed by its local CPU, and $t_k$ is the maximum tolerable latency. In this paper, it is assumed that the maximum tolerable latency for user $k$, $t_k$ is no longer than the channel coherence time, such that the wireless channels remain constant during a time slot with length $T$, i.e., $t_k\leq T, \forall k$, but can vary from time to time. The local CPU frequency of user $k$ is characterized by $f_k$, and $f_{k,m}$ is the computational speed of the edge cloud assigned to user $k$, where both of them are measured by the number of CPU cycles per second. Herein, a practical constraint that the total computing resources allocated to all the associated users must not excess the server’s computing capacity $F$, is given by $\sum_{k\in\mathcal{K}}f_{k,m}\leq F$.

In the following, the time latency and the energy consumption of user $k$ for our considered system are given in details.
\subsection{Latency}
\subsubsection{Local Computing at Users} 
Consider the local computing for executing the residual $(1-\lambda_k)R_k$ input bits  at user $k$, the time consumption for local computing at user $k$ is
\begin{equation}\label{local time}
t_{k,l}\!=\!\frac{c_{k}\left(1-\lambda_{k}\right)R_{k}}{f_{k}}.
\end{equation}
\subsubsection{Computation Offloading} 
According to the OFDMA mechanism, the inter-interference is ignored in virtue of the exclusive subcarrier allocation. Therefore, the aggregated transmission rate to offload $\lambda_kR_k$ input bits from user $k$ to MEC server is expressed as
\begin{equation}\label{r_k}
    r_k\!=\!B\sum_{n\in\mathcal{N}}x_{k,n}\log_2\left(1+\frac{p_{k,n}g_{k,n}}{\sigma_n^2}\right),
\end{equation}
where $g_{k,n}$ and $\sigma_n^2$ are the channel gain between user $k$ and BS, and the variance of the additive white Gaussian noise at BS on subcarrier $n$, respectively, where we set $\sigma_n^2=\sigma^2, \forall n$. Denote $p_{k,n}$ as the transmission power of user $k$ on subcarrier $n$. Apparently, any power optimization solution
has a good influence on the system performance. For the sake
of simplicity, we set $p_{k,n}=p_{k}^\text{max}/N_k$, where $p_{k}^\text{max}$ is the maximum transmission power and $N_k$ denotes the number of subcarriers allocated to user $k$.
Moreover, if user $k$ does not offload data to the MEC server, $p_{k,n} = 0$. Meanwhile, denote $x_{k,n}$ as the channel allocation indicator, specifically $x_{k,n}=1$ means that subcarrier $n$ is assigned to user $k$, otherwise $x_{k,n}=0$.


The offloading time $t_{k,\text{off}}$ of user $k$ mainly consists of two parts: the uplink transmission time $t_{k,u}$ from user $k$ to MEC server and the corresponding execution time at MEC server $t_{k,m}$. Therefore, the offloading time $t_{k,\text{off}}$ is given by
\begin{equation}\label{t_off}
\setlength{\abovedisplayskip}{3pt}
\setlength{\belowdisplayskip}{3pt}
t_{k,\text{off}}\!=\!t_{k,u}+t_{k,m}\!=\!\frac{\lambda_{k}R_{k}}{r_{k}}+\frac{\lambda_{k}R_{k}c_{k}}{f_{k,m}}.
\end{equation}
Due to the parallel computing at users and MEC server, the total latency for user $k$ depends on the larger one between $t_{k,l}$ and $t_{k,\text{off}}$, and can be expressed as $t_k\!=\!\max \{t_{k,l}, t_{k,\text{off}}\}$.

\subsection{Energy Consumption}
According to the strategy of computation offloading, the total energy consumption comprises two parts: the energy for local computing and for offloading, given in details as follow. 
\subsubsection{Local Computing mode}
Given the processor's computing speed $f_k$, the power consumption of the processor is modeled as $\kappa_kf_k^3$ (joule per second), where $\kappa_k$ represents the computation energy efficiency coefficient related to the processor's chip of user $k$ \cite{Bi2018Computation}. Taking consideration of \eqref{local time}, the energy consumption at this mode is given by
\begin{equation}
\setlength{\abovedisplayskip}{3pt}
\setlength{\belowdisplayskip}{3pt}
E_{k,l}\!=\!\kappa_kf_k^3t_{k,l}\!=\!\kappa_kc_k\left(1-\lambda_k\right) R_kf_k^2.
\end{equation}
\subsubsection{Computation offloading mode}
In this mode, the energy consumption includes the cost of uplink transmitting ($E_{k,u}$) and remote computing for offloaded $\lambda_kR_k$ input bits ($E_{k,m}$), which can be obtained as
\begin{equation}\label{e_u,m}
 E_{k,\text{off}}= \sum_{n\in\mathcal{N}}x_{k,n}p_{k,n}\frac{\lambda_kR_k}{r_k}+\kappa_m\lambda_kc_kR_kf_{k,m}^2,
\end{equation}
where $\kappa_m$ is the computation energy efficiency coefficient related to the processor's chip of MEC server. 

Therefore, the total energy consumption for user $k$ related with its computation offloading strategy in our system is $E_k=E_{k,l}+E_{k,u}+E_{k,m}$.

In this paper, we minimize the overall energy consumption of the considered system as follows
\begin{subequations}\label{OP1}
\begin{align}
\mathbf{P}:~\min _{\boldsymbol{\lambda},\boldsymbol{f},\boldsymbol{X}}~&\sum_{k\in\mathcal{K}}E_{k}\\
\mathrm{s.t.}  ~& 0\leq \lambda_{k}\leq 1,\forall k, \label{OP1-C1}\\
~&  \max \{t_{k,l}, t_{k,\text{off}}\} \leq T,\forall k,\label{OP1-C2}\\
~&  0 \leq f_{k,m}, \forall k,\label{OP1-C3}\\
~& \sum_{k\in\mathcal{K}}f_{k,m}\leq F,\label{OP1-C4}\\
~& \sum_{k\in\mathcal{K}}x_{k, n} \leq 1, \forall n,\label{OP1-C5}\\
~&  x_{k, n} \in\{0,1\},\forall k, n,\label{OP1-C6}
\end{align}
\end{subequations}
which is related to resource allocation on subcarriers, offloading communication and computation, and $\boldsymbol{\lambda}\triangleq\{\lambda_{k}\}$, $\boldsymbol{f}\triangleq\{f_{k,m}\}$ and $\boldsymbol{X}\triangleq\left\{x_{k,n}\right\}$. The constraints above can be explained as follows: constraint \eqref{OP1-C2} states that the task of user $k$ must be completely executed within a time slot; constraint \eqref{OP1-C3} and \eqref{OP1-C4} show that MEC server must allocate a positive computing resource to the user associated with it, and the sum of which cannot exceed the total computational capability of MEC server; constraint \eqref{OP1-C5} and \eqref{OP1-C6} enforce that each subcarrier can only be used by one user to avoid the multi-user interference.
\section{Proposed Algorithm}
In this section, we provide offloading and resource allocation strategy for the considered optimization problem  $\mathbf{P}$, which is intractable to deal with due to the coupled variants in both the constraints and the objective function based on our observation. To decouple these variants, we will divide the orignal problem $\mathbf{P}$ into two subprolems: 1) $\mathbf{P_O}$, offloading ratio selection; 2) $\mathbf{P_S}$, subcarriers and computing resource allocation. 

\textbf{Firstly}, with given computation capability assignment $\boldsymbol{f}$ and subcarrier allocation strategy $\boldsymbol{X}$, we can obtain the optimal offloading ratio $\boldsymbol{\lambda}^{\star}$ at the outer loop. \textbf{Secondly}, with the newly obtained offloading ratio $\boldsymbol{\lambda}^{\star}$, we can optimize $(\boldsymbol{f},\boldsymbol{X})$ at one iteration at its inner loop, and renew auxiliary variable $\phi$. Then, with the newly optimized $(\boldsymbol{f},\boldsymbol{X})$, we further update the dual variables $(\boldsymbol\alpha,\boldsymbol\beta,\gamma)$ at the next iteration at its inner loop.
\textbf{Finally}, we will iteratively update the derived $(\boldsymbol{\lambda},\boldsymbol{f},\boldsymbol{X})$ at the outer loop, and the procedures are known as the BCD method \cite{Richt2014Iteration,zhao2017exploiting}.
In this section, the joint optimization on offloading ratio, and subcarriers and computing resource allocation will be proposed in accordance with the iterative approach based on BCD method as follows.
\subsection{Offloading ration selection}\label{3B}
Given computation capability assignment $\boldsymbol{f}$ and subcarrier allocation strategy $\boldsymbol{X}$, the optimal offloading ratio $\boldsymbol{\lambda}^{\star}$ can be obtained by solving the following problem, 
\begin{equation}\label{OP1l}
\begin{aligned}
\mathbf{P_O}:~\min _{\boldsymbol{\lambda}}~&\sum_{k\in\mathcal{K}}E_{k}\\
\mathrm{s.t.}  ~&\eqref{OP1-C1}\eqref{OP1-C2},
\end{aligned}
\end{equation}
which can be decoupled into $K$ subproblems for each user, given by
\begin{subequations}\label{OP1lA}
\begin{align}
\mathbf{P_{O}1}:~\min _{\lambda_{k}}~&E_{k}\\
\mathrm{s.t.}  ~& 0\leq \lambda_{k}\leq 1,\label{OP1lA-C1}\\
~&  1-\frac{Tf_{k}}{c_{k}R_{k}} \leq \lambda_{k} \leq \frac{Tr_{k}f_{k,m}}{R_{k}f_{k,m}+r_{k}R_{k}c_{k}},\label{OP1lA-C2}
\end{align}
\end{subequations}
where \eqref{OP1lA-C2} can be derived with the help of \eqref{local time} and \eqref{t_off}. Apparently, $\mathbf{P_{O}1}$ is a convex problem with respect to (w.r.t.) $\lambda_k$, and the optimal offloading ratio $\lambda_k^\star$ at user $k$ can be achieved according to the following theorem.
\begin{theorem}\label{theorem1}
Given $(\boldsymbol{f},\boldsymbol{X})$, the optimal $\lambda_k^\star$ for $\mathbf{P_{O}1}$ is
\begin{equation}\label{OP-lkm}
\lambda_{k}^\star= 
\left\{
\begin{aligned}
	&\max\left\{1-\frac{Tf_{k}}{c_kR_{k}}, 0\right\}, &&\text{if}~\frac{\partial E_{k}}{\partial \lambda_{k}} \geq 0,\\
	&\min \left\{\frac{Tr_kf_{k,m}}{R_kf_{k,m}+r_kR_kc_{k}},1\right\}, &&\text{otherwise}.
\end{aligned}
\right.
\end{equation}
\end{theorem}
\begin{IEEEproof}
It can be obtained resorting to the first-order condition and comparing with the boundaries points provided by \eqref{OP1lA-C1} and \eqref{OP1lA-C2}.
\end{IEEEproof}

\subsection{Subcarriers and computing resource allocation strategy}\label{3A}
With the newly obtained offloading ratio $\boldsymbol{\lambda}^\star$, the resource allocation strategy will be designed to assign the computation capability of MEC server, and allocate the subcarriers for each user, to further reduce the energy consumption. In this subsection, we aim to minimize the energy consumption for computation offloading mode via the following optimization problem, the objective function of which is $\sum_{k\in\mathcal{K}}E_{k,\text{off}}$ indeed,
\begin{subequations}\label{OP2}
	\begin{align}
	\mathbf{P_S}:~\min _{\boldsymbol{f},\boldsymbol{X}}~&\sum_{k\in\mathcal{K}}\sum_{n\in\mathcal{N}}x_{k,n}p_{k,n}\frac{\lambda_kR_k}{r_k}\!+\!\kappa_m\lambda_kc_kR_kf_{k,m}^2\\
	\mathrm{s.t.}
	~& \eqref{OP1-C3}-\eqref{OP1-C6}.\nonumber\\
	~& t_{k,u}+t_{k,m}\leq T, \forall k,\label{OP2-CT}
	\end{align}
\end{subequations}
where the time-sensitive constraint \eqref{OP1-C2} can be recast as \eqref{OP2-CT} when $\boldsymbol{\lambda}$ is given.

However, we cannot transform the primal domain of $\mathbf{P_S}$ into the dual domain directly since $r_k$ is in the denominator and its form in \eqref{r_k} makes the problem more intractable. Therefore, a new non-negative auxiliary variable $\boldsymbol{\phi}\triangleq\{\phi_{k}\}$ will be introduced to transform $\mathbf{P_S}$ into the following problem,
\begin{subequations}\label{OP2-C}
	\begin{align}
	\mathbf{P_S1}:~\min _{\boldsymbol{f},\boldsymbol{X},\boldsymbol{\phi}}~&\sum_{k\in\mathcal{K}}\sum_{n\in\mathcal{N}}x_{k,n}p_{k,n}\frac{\lambda_kR_k}{\phi_k}+\kappa_m\lambda_kc_kR_kf_{k,m}^2\\
	\mathrm{s.t.}
	~&\eqref{OP1-C1}-\eqref{OP1-C6},\eqref{OP2-CT}, \nonumber\\
	~& 0\leq\phi_{k}\leq r_{k}, \forall k, \label{OP2-C1}
	\end{align}
\end{subequations}
where we can update $\boldsymbol{\phi}$ to help minimize the energy consumption. The Lagrangian for the above problem with the given $\boldsymbol{\lambda}^\star$ can be written as
\begin{align}\label{OP2-L}
  &\!\mathcal{L}(\boldsymbol{f},\boldsymbol{X},\boldsymbol\phi,\boldsymbol\alpha,\boldsymbol\beta,\gamma)=\sum_{k\in\mathcal{K}}\sum_{n\in\mathcal{N}}x_{k,n}p_{k,n}\frac{\lambda_kR_k}{\phi_k}\nonumber\\
  &~~+\sum_{k\in\mathcal{K}}\kappa_m\lambda_kc_kR_kf_{k,m}^2+\sum_{k\in\mathcal{K}}\alpha_{k}\left(\frac{\lambda_kR_{k}}{\phi_{k}}+\frac{\lambda_kR_kc_{k}}{f_{k,m}}-T\right)\nonumber\\
  &~~+\sum_{k\in\mathcal{K}}\beta_{k}(\phi_k-r_{k})+\gamma\!\left(\sum_{k\in\mathcal{K}}f_{k,m}-F\right),
  \end{align}
  where 
  $\boldsymbol\alpha\triangleq\{\alpha_k\}$, $\boldsymbol\beta\triangleq\{\beta_k\}$, and $\gamma$ are the non-negative Lagrange multipliers corresponding to the related contraints. Define $\mathcal{F}$ as all sets of possible $\boldsymbol{f}$ that satisfy constraint \eqref{OP1-C3}, $\mathcal{X}$ as all sets of possible $\boldsymbol{X}$ that satisfy constraints \eqref{OP1-C5} and \eqref{OP1-C6},
  The Lagrange dual function is then defined as
  \begin{equation}\label{OP2-DF}
  \begin{aligned}
  \mathbf{g}(\boldsymbol\alpha,\boldsymbol\beta,\gamma)\!=\!\min_{\boldsymbol{f}\in\mathcal{F},\boldsymbol{X}\in \mathcal{X},\boldsymbol{\phi}\in\mathcal{Q}}\mathcal{L}(\boldsymbol{f},\boldsymbol{X},\boldsymbol\phi,\boldsymbol\alpha,\boldsymbol\beta,\gamma).
  \end{aligned}    
  \end{equation}
  Furthermore, the Lagrange dual problem is given by
  \begin{equation}\label{OP2-DP}
  \begin{aligned}
  \max ~&\mathbf{g}(\boldsymbol\alpha,\boldsymbol\beta,\gamma)\\
  \mathrm{s.t.} ~&\boldsymbol{\alpha}\succeq \boldsymbol{0},~\boldsymbol{\beta}\succeq \boldsymbol{0},
  ~\gamma \geq 0.
  \end{aligned}    
  \end{equation}
  
  With the given offloading ratio and transmission power, the following steps for updating are adopted to obtain the optimal solutions for computing resource and subcarrier allocation.
\subsubsection{Computational capabilities assignment}
Employing the KKT conditions, the following condition is both necessary and sufficient for computation capability assignment's optimality:
\begin{align}\label{OP3-fkm}
\setlength{\abovedisplayskip}{3pt}
\setlength{\belowdisplayskip}{3pt}
\frac{\partial\mathcal{L}}{\partial{f_{k,m}}}= 2f_{k,m}\kappa_m\lambda_kR_kc_{k}-\frac{\alpha_kc_kR_{k}\lambda_{k}}{f_{k,m}^2}+\gamma=0.
\end{align}
However, it is difficult to find a closed-form expression for the optimal solution, $f_{k,m}^\star$. Fortunately, we can resort to the following proposition to obtain $f_{k,m}^\star$.

Since $\mathcal{L}$ is a convex function of $f_{k,m}$, and $\frac{\partial\mathcal{L}}{\partial{f_{k,m}}}$ increases monotonically along with $f_{k,m}$, we can adopt the bisection method to obtain $f_{k,m}^\star$ within $0\leq f_{k,m}\leq F$. The detailed process of achieving $f_{k,m}^\star$ is summarized in Algorithm 1.
\begin{algorithm}[h]  
	\caption{Proposed Binary Search Algorithm}  
	{\bf Input:} Given offloading ratio $\boldsymbol\lambda^\star$, accuracy indicator $\epsilon$ and transmission power $\boldsymbol{p}$.
	\begin{algorithmic}[1]
		\For {$k\in\mathcal{K}$}
		\State {\bf Initialize:} $f_{k,m}^\text{UB} = F$ {\bf{and}} $f_{k,m}^\text{LB}=0$;
		\Repeat
		\State Set $f_{k,m}$ = $\frac{1}{2}(f_{k,m}^\text{UB}+f_{k,m}^\text{LB})$;
		\State Compute $\frac{\partial{L}}{\partial{f_{k,m}}}$ according to \eqref{OP3-fkm};
		\If{$\frac{\partial{\mathcal{L}}}{\partial{f_{k,m}}}>0$}
		\State set $f_{k,m}^\text{UB} = f_{k,m}$;
		\Else
		\State set $f_{k,m}^\text{LB} = f_{k,m}$;
		\EndIf 
		\Until {$\left|\frac{\partial \mathcal{L}}{\partial f_{k,m}}\right|\leq\epsilon_3$;}
		\State Obtain the optimal $f_{k,m}^\star$.
		\EndFor
	\end{algorithmic}  
\end{algorithm}
\subsubsection{Subcarrier allocation strategy}\label{SAS}
With the achieved optimal computing capability assignment, the optimal subcarrier allocation can be obtained through the following procedures. With some mathematic manipulations, we can rewrite \eqref{OP2-L} as
\begin{align}\label{OP2-L2}
&\!\mathcal{L}(\boldsymbol{X},\boldsymbol\phi,\boldsymbol\alpha,\boldsymbol\beta,\gamma)
=\sum_{k\in\mathcal{K}}\sum_{n\in\mathcal{N}}x_{k,n}\bigg[p_{k,n}\frac{\lambda_kR_k}{\phi_k}\nonumber\\
&-B\beta_k\log_2(1\!+\!p_{k,n}\tilde{g}_{k,n})\bigg]\!+\!\sum_{k\in\mathcal{K}}\omega_{k}\!-\!\gamma F,
\end{align}
where 
\begin{equation}\label{OP3-w}
\begin{aligned}
\omega_{k}~&=\alpha_{k}\!\left(\!\frac{\lambda_kR_{k}}{\phi_{k}}\!+\!\frac{\lambda_kR_kc_{k}}{f_{k,m}}\!-\!T\right)+\gamma f_{k,m}\!+\!\beta_{k}\phi_k\\
~&+\kappa_m\lambda_kc_kR_kf_{k,m}^2.
\end{aligned}
\end{equation}
On the observation of \eqref{OP2-L2}, we further suppose that subcarrier $n$ is assigned to user $k$, we have
\begin{equation}\label{OP2-L3}
\begin{aligned}
~&\mathcal{L}\!=\! \sum_{n\in\mathcal{N}}\mathcal{L}_{n}\!+\!\sum_{k\in\mathcal{K}}\omega_{k}\!-\!\gamma F,
\end{aligned}    
\end{equation}
where 
\begin{equation}\label{OP2-Ln1}
\begin{aligned}
~\mathcal{L}_{n}\!=\!p_{k,n}\frac{\lambda_kR_k}{\phi_k}\!-\!B\beta_k\log_2(1\!+\!p_{k,n}\tilde{g}_{k,n}).
\end{aligned}    
\end{equation}
Thus, the subproblem is given by
\begin{equation}\label{OP2-SUB}
\begin{aligned}
\min_{\boldsymbol{X}_{n}\in\mathcal{X}}\mathcal{L}_{n}(\boldsymbol{X}_{n},\boldsymbol{\phi},\boldsymbol{\beta}),
\end{aligned}    
\end{equation}
where $\boldsymbol x_{n}\!=\!\left\{x_{k, n}\right\}_{k=1}^{K}$, and this problem can be solved independently.
To minimize each $\mathcal{L}_{n}$, the optimal $\boldsymbol{x}_n$
can be obtained as
\begin{equation}{\label{OP2-Xn}}
x_{k,n}^\star=\left\{\begin{array}{l}{1,~\text {if}~k=k^\star=\operatorname{argmin}_{k}~\mathcal{L}_{n}}, \\ {0,
	~\text{otherwise}.}\end{array}\right.
\end{equation}
\subsubsection{Auxiliary variable $\boldsymbol{\phi}^\star$ selection}
In this part, with the newly obtained $\{\boldsymbol{f}^\star,\boldsymbol{X}^\star\}$ in the above subsections, we now try to find the optimal $\boldsymbol{\phi}^\star$, which can be solved via the following optimization problem,
\begin{subequations}\label{OP3-phi}
\begin{align}
\mathbf{P_S2}:\min_{\boldsymbol{\phi}}&\sum_{k\in\mathcal{K}}\left[\sum_{n\in\mathcal{N}_k}\frac{p_{k,n}\lambda_{k}R_{k}}{\phi_k}\!+\!\frac{\alpha_{k}\lambda_{k}R_{k}}{\phi_{k}}\!+\!\beta_{k}\phi_{k}\right]\\
\mathrm{s.t.}~&  \eqref{OP2-CT},\nonumber\\
~& 0\leq\phi_{k}\leq \tilde{r}_k, \forall k, \label{OP3-Cphi2}
\end{align}
\end{subequations}
where $\tilde{r}_k\triangleq \sum_{n\in\mathcal{N}_k} B\log_{2}(1+p_{k,n}\tilde{g}_{k,n})$, and $\mathcal{N}_k$ denotes the set of subcarriers allocated to user $k$ through \eqref{OP2-Xn}. It is obvious that $\mathbf{P_S2}$ can be further decoupled into $K$ subproblems w.r.t. each user, given by
\begin{subequations}\label{OP3-phi-sub}
\begin{align}
\mathbf{P_S3}:\min_{\phi_k}&\left[\frac{\lambda_{k}R_{k}}{\phi_k}\sum_{n\in\mathcal{N}_k}p_{k,n}+\frac{\alpha_{k}\lambda_{k}R_{k}}{\phi_{k}}+\beta_{k}\phi_{k}\right]\\
\mathrm{s.t.}~& \eqref{OP3-Cphi2} \nonumber\\
~& \frac{\lambda_{k}R_{k}}{\phi_{k}} + \frac{\lambda_{k}R_{k}c_{k}}{f_{k,m}}\leq T,\label{OP3-phi-sub-C1}
\end{align}
\end{subequations}
the optimal auxiliary variable $\boldsymbol{\phi}^\star$ can be obtained by the following theorem.
\begin{theorem}\label{theorem-phi}
Given the optimal computation capability assignment $\boldsymbol{f}^\star$, and the optimal subcarrier allocation strategy $\boldsymbol{X}^\star$, the optimal auxiliary variable $\phi^{\star},\forall k$ is given by
\begin{equation}{\label{OP3-Xn2}}
\phi_{k}^\star\!=\!\left\{
\begin{aligned}
&{\phi_{k,1}, ~\text{if}~\phi_k^o <\phi_{k,1}},\\
&{\phi_k^o,~~~\text {if}~\phi_{k,1}\leq \phi_k^o\leq \tilde{r}_{k}}, \\ 
&{\tilde{r}_{k},
	~~~\text{otherwise}},
\end{aligned}\right.
\end{equation}
where $\phi_{k,1}$ and $\phi_k^o$ are given, respectively, by
\begin{eqnarray}
\phi_{k,1}&=&\frac{\lambda_{k}R_{k}f_{k,m}}{Tf_{k,m}-\lambda_{k}R_{k}c_{k}},\\
\phi_k^o&=&\sqrt{\left(\alpha_{k}+\sum_{n\in\mathcal{N}_k}p_{k,n}\right)\tilde{\lambda}_k},
\end{eqnarray}
where $\tilde{\lambda}_k=\frac{\lambda_kR_k}{\beta_k}$.
\end{theorem}
\begin{IEEEproof}
Due to the space limitation, the proof is omitted and presented in \cite{Zhao2020EnergyAwareOI}.
\end{IEEEproof}

\subsubsection{Lagrange Multipliers Update} In this subsection, since $\boldsymbol{f^\star}$,  $\boldsymbol{X^\star}$ and $\boldsymbol{\phi}^\star$ are obtained, we can deal with the dual problem \eqref{OP2-DP}, which is a convex function, by updating $\boldsymbol{\alpha}$, $\boldsymbol{\beta}$ and $\gamma$ using the subgradient method.
The dual variables $(\boldsymbol\alpha,\boldsymbol\beta,\boldsymbol\beta,\gamma)$ can be updated according to the following formulations,
\begin{equation}\label{OP3-UD}
\begin{aligned}
&\alpha_{k}^{z+1}=\left[\alpha_{k}^{z}-\zeta_{k}\left
(\frac{\lambda_kR_{k}}{\phi_{k}}+\frac{\lambda_kR_kc_{k}}{f_{k,m}}-T\right)\right]^{+},\\
&\beta_{k}^{z+1}=\left[\beta_{k}^{z}-\xi_{k}\left(\phi_k-r_{k}\right)\right]^{+},\\
&\gamma^{z+1}=\left[\gamma^{z}-\theta\left(\sum_{k\in\mathcal{K}}f_{k,m} - F\right)\right]^{+},
\end{aligned}
\end{equation}
where $\zeta_k$, $\xi_k$, and $\theta$ are the stepsizes related to each dual variable during iterations, given in TABLE \uppercase\expandafter{\romannumeral1}.
 \begin{table}  
  \caption{SIMULATION PARAMETERS}  
  \begin{tabularx}{9cm}{lXl}  
  \hline  
  \bf MEC System Parameters   &   \bf Values \\  
  \hline  
  The CPU frequency of MEC sever  &  10GHz \\
  The CPU frequency of mobile users & 0.6-0.7GHz\\
  The transmission power of users        & 27.8dBm\\
  Input data size of users          & $1000 - 1500$ bits\\
  Maximum accomplished deadline $T$   &2ms \\
  The computation workload/intensity $c_{k}$ & $1000-1200$ cycles/bit\\
  Background noise $\sigma^{2}$     &$10^{-13}$W\\
  Subcarrier bandwidth $B$            &12.5KHz\\
  \hline  
  \bf Lagrange Iteration Parameters   &   \bf Values \\  
  \hline  
  Maximum number of iterations $z_\text{max}$ &600\\
  Iterations precision & $10^{-5}$\\
  Stepsizes $\zeta_k$, $\eta_{k}$, $\theta$       & $10^{-18}$, $10^{-6}$, $10^{-5}$\\
  \hline  
  \end{tabularx}\vspace{-0.85em}
  \end{table}

According to \ref{3A} and \ref{3B}, the details to solve $\mathbf{P}$ are summarized in Algorithm 2 as follows.
\begin{algorithm}[h]
	\begin{algorithmic}[1]
		\caption{Proposed Algorithm}  
		\State {\bf Initialization:} Given $\{\boldsymbol{f},\boldsymbol{X},\boldsymbol{P}, \boldsymbol{\phi},\boldsymbol\alpha,\boldsymbol\beta,\gamma\}$, we set $z=0$, and denote $z_{\text{max}}$ as the maximum number of iterations.
		\Repeat (from 2 to 14)
		\State Determine offloading ratio $\boldsymbol\lambda$ according to \eqref{OP-lkm};
		\Repeat (from 4 to 12)
		\Repeat (from 5 to 10)
		\State Assign computing resource $\boldsymbol{f}$ by Algorithm 1; 
		\State Determine subcarrier allocation $\boldsymbol{X}$ via \eqref{OP2-Xn};
		\State Update $p_{k,n}=\frac{p_k^{max}}{N_k}$;
		\State Update $\boldsymbol\phi^{\star}$;
		\Until Lagrangian function converges;
		\State Update $\boldsymbol\alpha$, $\boldsymbol{\beta}$, and $\gamma$ resorting to \eqref{OP3-UD}; 
		\Until {$\boldsymbol\alpha, \boldsymbol\beta, \gamma$ converge;}
		\State $z = z + 1$;
		\Until $z>z_{\text{max}}$.
	\end{algorithmic}  
\end{algorithm}
\section{NUMERICAL RESULTS}
In the simulations, the path loss model is Rayleigh distributed and denoted by $|\beta|d^{-2}$, where $|\beta|$ and $d$ represent the short-term channel fading and the distance between two nodes, respectively. User devices have the same maximum transmit power, which are evenly and independently distributed in a circular area around the MEC server with a radius of 30 meters. Moreover, we set $\kappa_k$ and $\kappa_m$ as $10^{-24}$ and $10^{-26}$, respectively, and other parameters employed in the simulations are summarized in TABLE 1, unless otherwise mentioned.

This section presents the numerical results to demonstrate the better performance of our proposed algorithm compared with the conventional schemes: 1) \textbf {Local computation algorithm (LC)}, where all user devices process their own tasks by local CPU. 2) \textbf {Fixed offloading ratio algorithm (FR)} where we assign fixed offloading ratios for each user satisfying the time-sensitive constraints \eqref{OP1-C2}. 
\begin{figure}
    \centering
    \includegraphics[width=6.5cm]{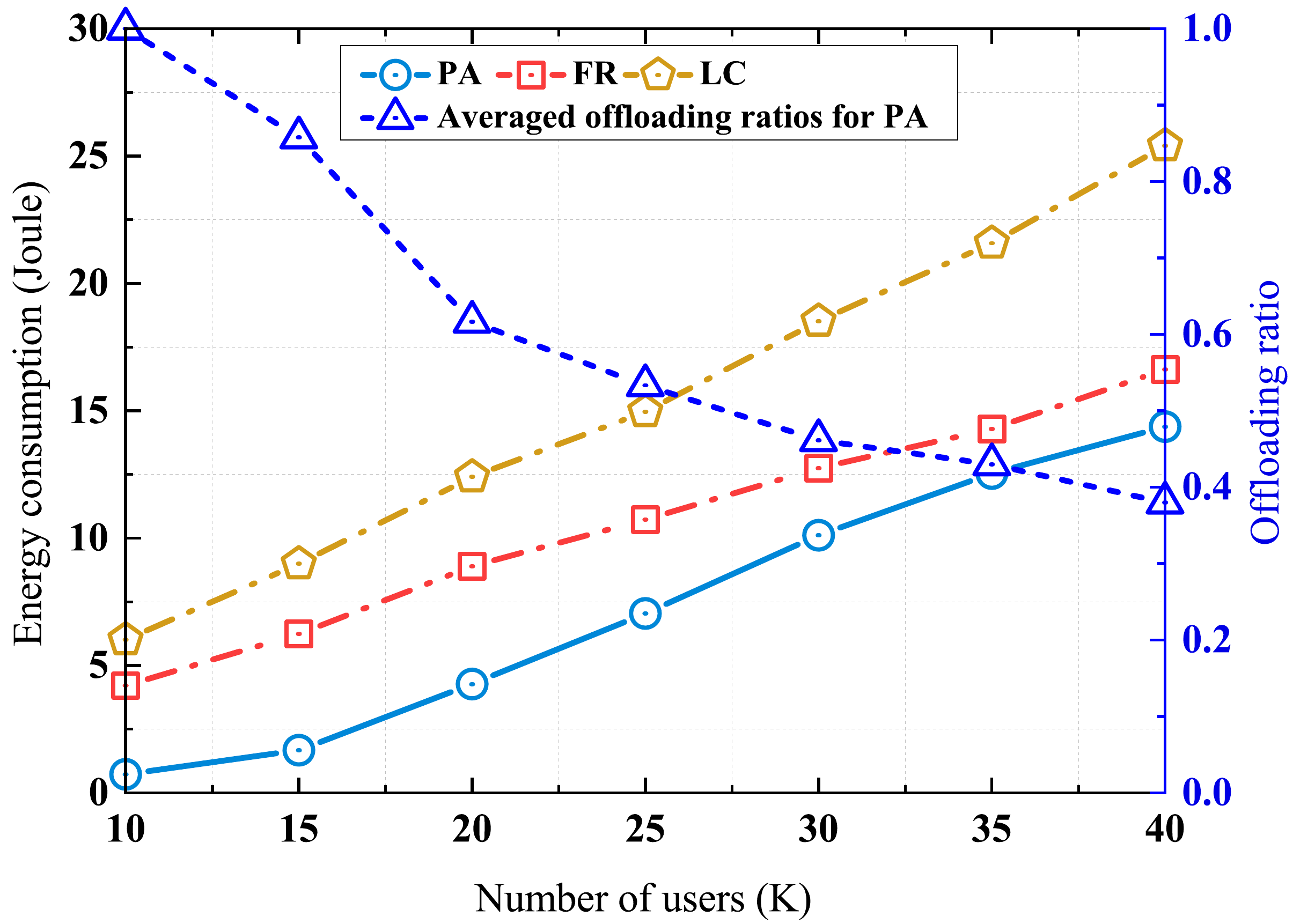}\\
    \caption{The comparison of the total energy consumption (and the averaged
    offloading ratios for PA where $K\!=\!20$) versus different number of users where $N\!=\!512$.}
    \label{fig1}
\end{figure}

In Fig.\ref{fig1}, for the proposed and reference schemes, we plot the total energy consumption w.r.t. different number of users. With the growth of the number of users, the total energy consumption of different algorithms is increasing, and our proposed algorithm (PA) can help save 40\% -- 70\% and 20\%-50\% energy consumption compared with LC and FR. Moreover, it can be seen that the averaged offloading ratio will decrease with the increase of the number of users, since computing resource allocated to each user by MEC is declining, and thus users will cut down the offloading ratio.

\begin{figure}
    \centering
    \includegraphics[width=6.5cm]{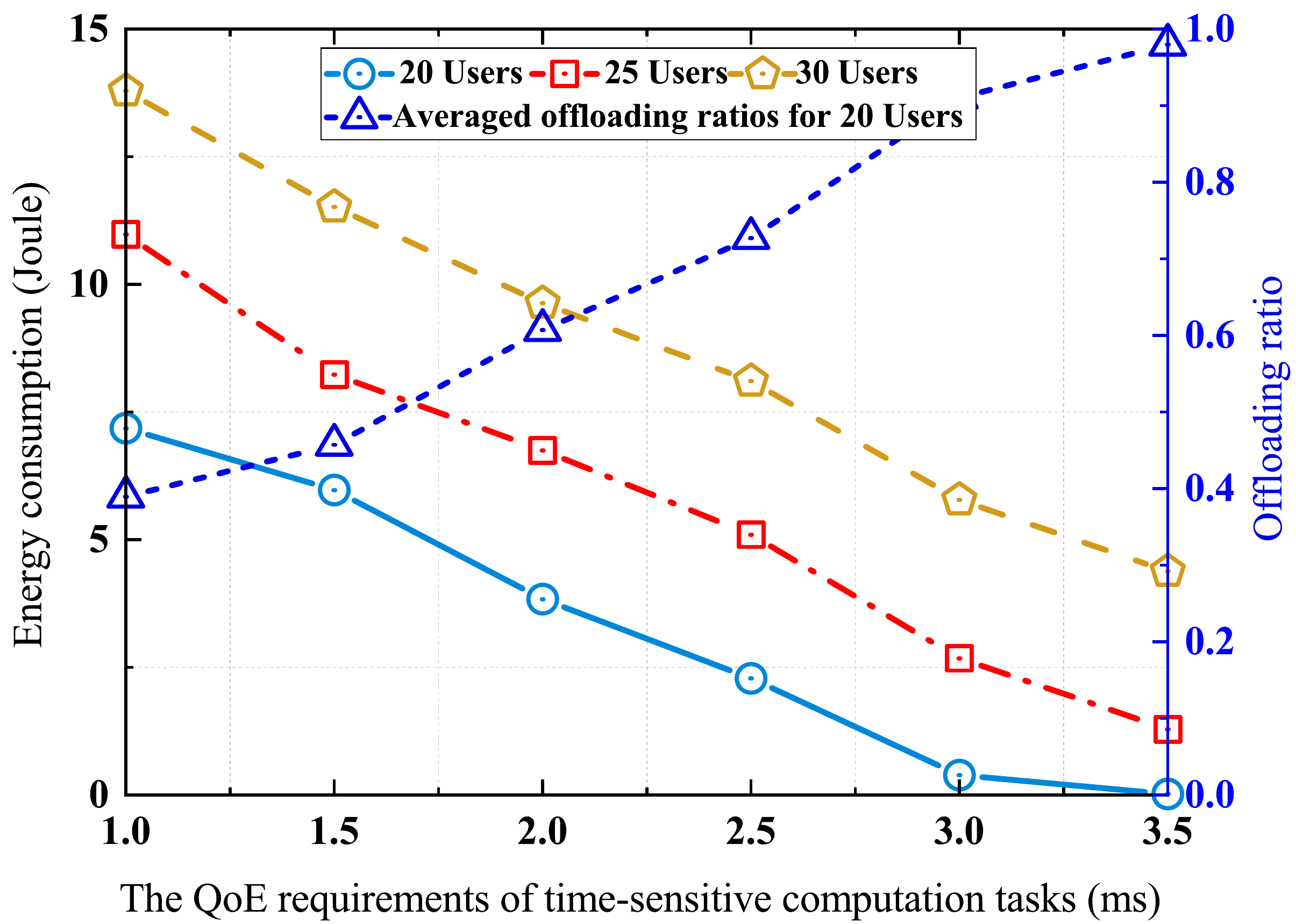}\\
    \caption{The comparisons of the total energy consumption (and the averaged
    offloading ratios where $K\!=\!20$) versus different QoE requirements
    of time-sensitive computation tasks of users where $N\!=\!512$.}
    \label{fig2}
\end{figure}

In Fig.\ref{fig2}, we study the influence of the QoE requirements of time-sensitive computation tasks of users on the total energy consumption for different numbers of users. On the observation of Fig.\ref{fig2}, the total energy consumption is reducing along with the gradually decreasing QoE requirement of time-sensitive computation tasks. This is because users can ask for more help from MEC server by uploading more data since the requirement of latency is not so strict shown by the blue line, and thus users can reduce more energy dissipation.

\begin{figure}
    \centering
    \includegraphics[width=6.5cm]{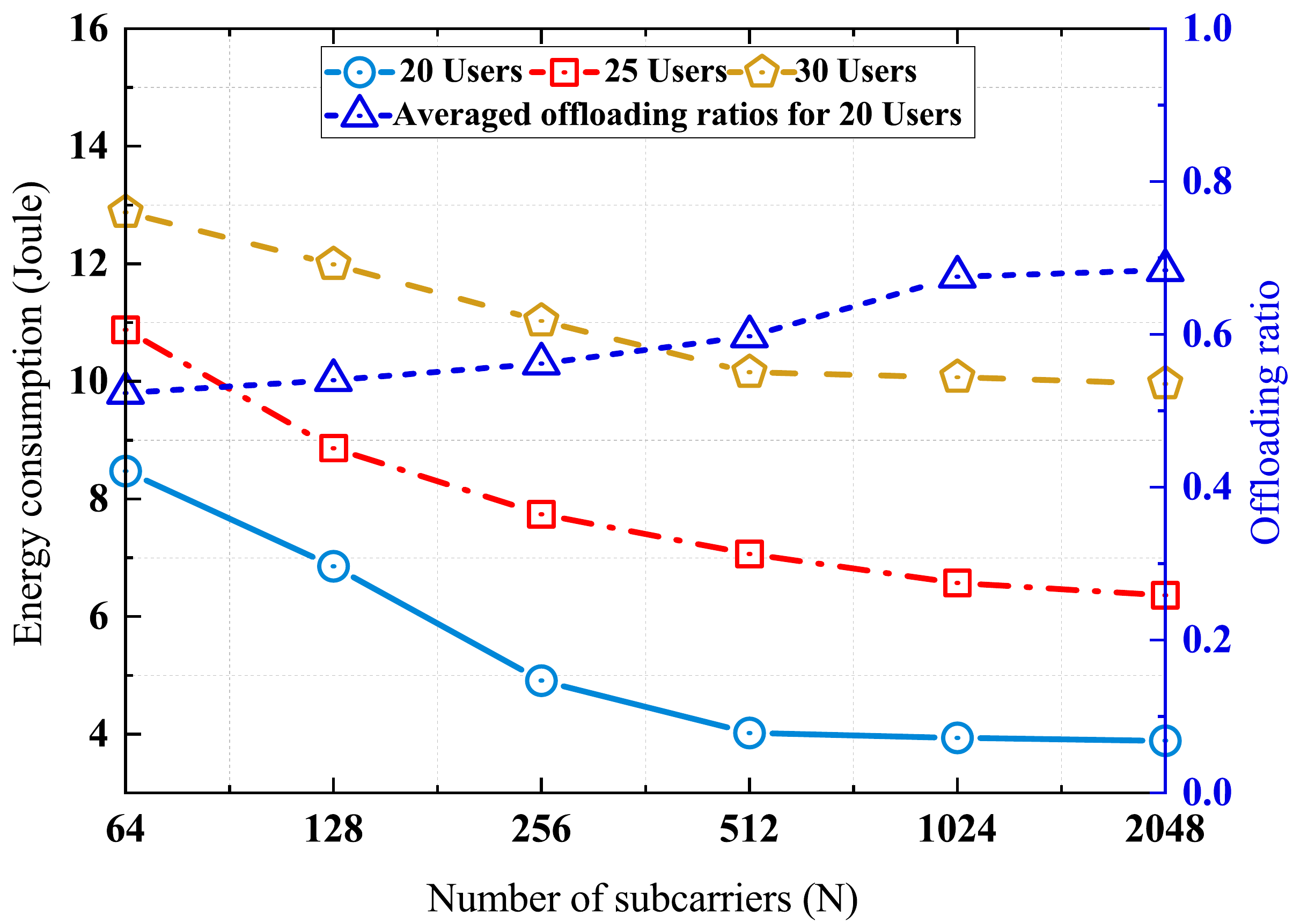}\\
    \caption{The comparisons of the total energy consumption versus different number of subcarriers where $K\!=\!20,25,30$.}
    \label{fig3}
\end{figure}

In Fig.\ref{fig3}, we evaluate the effect of the number of subcarriers on the total energy consumption. Moreover, it can be seen that the total energy consumption is reducing along with the increasing number of subcarriers. The reason is that the user will have a better chance to select the subcarrier with preferable channel gain to get the reduction of transmission power in turn, meanwhile, the user can offload more data with the same amount of transmission power shown by the blue line, and thus reduce the energy consumption with the aid of MEC server.
\section{CONCLUSIONS}
In this paper, the energy minimization was investigated by jointly optimizing partial offloading strategy and resource allocation for the OFDMA-based MEC networks, while satisfying the QoE requirements of time-sensitive computation tasks for all users. Due to the coupled optimization variables,  we decomposed the formulated minimization problem into two subproblems, and optimized them sequentially instead of solving the original problem directly. Simulation results demonstrate that the proposed algorithm can effectively reduce the total energy consumption of the network, and outperform the reference schemes with a better performance.
\bibliographystyle{IEEEtran/bibtex/IEEEtran}
\bibliography{main.bib}
\end{document}